\documentstyle[twoside,fleqn,espcrc2,epsf]{article}
\newcommand{\msbar}{{\rm \overline{MS\kern-0.14em}\kern0.14em}}

\begin{document}

% Liverpool Preprint: LTH 289\\
%October 6, 1992

\title{The Running Coupling from SU(3) Potentials}

\author{ UKQCD Collaboration - presented by Chris Michael
\address{DAMTP, University of Liverpool, Liverpool L69~3BX, UK}}

\begin{abstract}

 From an accurate determination of the inter-quark potential, one can
study the running coupling constant for a range of $R$-values
and hence estimate the scale $\Lambda_{\msbar} $.
Detailed results are  presented  for $SU(3)$  pure  gauge  theory
from a study of a $36^4$ lattice at $\beta=6.5$.

\end{abstract}

\maketitle

\section{Introduction}

In the continuum the potential between static quarks is known
perturbatively to two loops in terms of  the  scale $\Lambda_{\msbar} $.
 For  $SU(3)$ colour, the continuum force is given by \cite{bill}

\begin{equation}
{dV \over dR } =  {4 \over 3} {\alpha(R) \over R^2}
\end{equation}

\noindent with the effective coupling $\alpha (R)$ given by

\begin{equation}
 { 1 \over 4 \pi [ b_0  \log (R\Lambda _R )^{-2} +
(b_1 / b_0 ) \log \log (R\Lambda_R )^{-2} ] }
\end{equation}

\noindent where $b_0=11/16 \pi ^2$ and $b_1=102 \ b_0^2/121$ are
 the usual coefficients in the
perturbative expression for the $\beta$-function
and, neglecting quark loops in the vacuum,  $\Lambda_R= 1.048
\Lambda_{\msbar}$.
Note that the usual lattice regularisation scale $\Lambda_L = 0.03471
 \Lambda_{\msbar}$.

 At large separation $R$, the potential behaves as $KR$ where $K$  is  the
string tension.  Thus in principle knowledge of  the  potential $V(R)$
 serves  to
determine the dimensionless ratio $\sqrt K/\Lambda $ which relates the
perturbative scale $\Lambda$ to a non-perturbative observable such as the
string tension $K$.  This is the  basis  of
the method~\cite{cmlms} used for $SU(2)$ which we extend here to $SU(3)$
colour.

For $SU(3)$ as for $SU(2)$, the bare lattice coupling proves to be a
poor guide to physical behaviour in that asymptotic scaling to two loops
is not yet manifest. However, the weaker requirement of scaling is well
satisfied:  the dimensionless ratios of physical quantities are
found to be independent of $\beta $.  For example in $SU(2)$, the
potential $V(R)$ scales
\cite{ukqcd} over a range of lattice spacing of a factor of 4 (from
$\beta =2.4$ to $ 2.85$).  That scaling but not asymptotic scaling is
valid implies that the bare coupling constant derived from $\beta $ is
inappropriate and that an effective coupling constant derived from some
physical quantity is a better choice.  This has been emphasized by
Lepage and Mackenzie \cite{lm}.  It is also the basis of the method
proposed by L\"uscher et al.~\cite{lu} to extract the running coupling
constant.

Here we choose to determine the running coupling constant from the
interquark potential between static quarks at small distance $R$.  This
quantity can be determined in a straightforward way from lattice
simulation on large volume lattices.  Although we require small $R$ and hence
large energy $1/R$ to make most precise contact with the perturbative
expression, the lattice method implies the presence of lattice
artefacts when $R \approx a$. Thus we need to work on the largest
spatial lattice available, consistent with avoiding finite size effects.
We present results from a $36^4$ lattice at $\beta=6.5$ to achieve
this~\cite{lth285}.
 These results are compared with previous UKQCD data~\cite{uksu3}
from $24^3\times 48$ lattices at $\beta=6.2$ to check scaling.

\section{Lattice potentials}

Previous studies of glueballs and potentials in quenched $SU(3)$
have found evidence~\cite{mt}
for finite size effects when the spatial lattice
size $L$ is such that $mL < 9$ where $m$ is the $0^{++}$ glueball mass.
This corresponds to $\sqrt{K} L < 2.6$. Our previous study on a
$24^3\times 48$ lattice~\cite{uksu3} at $\beta=6.2$ corresponded to
$\sqrt{K} L = 3.8$ which appears to be comfortably in the large
volume region where finite size effects are negligible. For our
study at larger $\beta$ we are constrained by the memory
limitation to a $36^4$ lattice and we selected $\beta=6.5$
since this again corresponds to the same spatial size in physical
units, namely $\sqrt{K} L=3.8$. Even though direct study of the
potential seems to suggest~\cite{bali} that it is relatively
insensitive to finite size effects, we prefer to adopt the cautious
policy of conducting our study at a fixed large physical volume.

The details of our analysis of the potential between static sources
on these $36^4$ lattices are given elsewhere~\cite{lth285}. Here
we summarise the main method and results. We conducted a relatively
high statistics study of the potential at large $R$ to be able
to get an accurate value of the string tension $K$. From a fit
to $R$-values from $4a$ to $24a$ we obtain $Ka^2=0.0114(4)$ where
a systematic error from $T$-extrapolation is included. The values
of the force are collected in table~\ref{force}.

In order to explore the potential at small $R$, we measure at
off-axis as well as on-axis separations. This will allow us to
measure the extent to which rotational invariance is broken down
to the cubic invariance on the lattice. Furthermore, we find that
a simple parametrisation in terms of a lattice one gluon exchange
is able to parametrise successfully our results. This is
non-trivial since there is only one free parameter to accommodate
the breaking of rotational invariance and yet a satisfactory fit
is obtained to all the off-axis and on-axis data. The details of
the fitting procedure are exactly those pioneered successfully for
$SU(2)$~\cite{cmlms}.

{}From this successful parametrisation of the departure from
rotational invariance, we can construct a corrected potential
by replacing the lattice gluon contribution in the fit by the
full continuum gluon propagator. These results are shown in
table~\ref{alphat}.

\par
\begin{table}
\begin{center}
\caption{ \label{force}
 The force $\Delta V/\Delta R$ at
average separation $R$ derived from $T$-ratio 4/3.
}
\begin{tabular}{ r l }\hline
 $R/a$ \  & $\ \Delta V/\Delta R $    \\\hline
 2.4142 & 0.0667(4)   \\
 3.4142 & 0.0344(4)   \\
 4.2361 & 0.0286(8)   \\
 5.0645 & 0.0226(5)  \\
 5.8284 & 0.0196(19)  \\
 6.1623 & 0.0190(16)  \\
 6.7678 & 0.0174(9)   \\
 7.6056 & 0.0161(10)  \\
 8.2426 & 0.0161(11)  \\
 9.0000 & 0.0149(3)   \\
11.0000 & 0.0130(4)   \\
13.0000 & 0.0134(5)   \\
15.0000 & 0.0121(5)   \\
17.0000 & 0.0132(6)   \\
19.0000 & 0.0117(6)   \\
21.0000 & 0.0125(8)   \\
23.0000 & 0.0126(7)     \\\hline
\end{tabular}
\end{center}
\end{table}

\begin{table}
\begin{center}
\caption{ \label{alphat}
The force $\Delta V/\Delta R$ and lattice artefact corrected force
$ \Delta V_c / \Delta R $ at
average separation $R$. The running coupling $\alpha (R)$ derived
from the corrected force is shown as well. The second error shown on
$\alpha$ is $10\%$ of the lattice artefact correction.
}
\begin{tabular}{ r l c l }\hline
 $R/a$ \  & $\ \Delta V/\Delta R $  & $\Delta V_c / \Delta R$ &
 $ \qquad \alpha(R)$  \\\hline
 1.2071 & 0.2067(7)    & 0.1607    & 0.170(1)(5)    \\
 1.7071 & 0.0750(7)    & 0.0930    & 0.197(1)(4)    \\
 2.1180 & 0.0959(19)   & 0.0664    & 0.223(6)(10)   \\
 2.5322 & 0.0541(5)    & 0.0523    & 0.248(2)(1)   \\
 2.9142 & 0.0263(40)   & 0.0424    & 0.270(26)(10)  \\
 3.0811 & 0.0471(39)   & 0.0368    & 0.262(28)(7)  \\
 3.3839 & 0.0391(12)   & 0.0371    & 0.317(10)(2)   \\
 3.9241 & 0.0292(5)    & 0.0290    & 0.333(6)(1)  \\
 \hline
\end{tabular}
\end{center}
\end{table}

\section{Running Coupling}

It is now  straightforward  to  extract  the  running  coupling
constant by using

\begin{equation}
\alpha( { R_1 + R_2 \over 2 }) = { 3\over 4} R_1 R_2 { V_c(R_1)-V_c(R_2) \over
 R_1-R_2 }
\end{equation}

\noindent where the error in using a finite difference is here negligible.
This is shown in table~\ref{alphat} and is plotted  in the figure versus
$R \sqrt K$ where $K$ is taken from the large $R$ fit.
The interpretation of $\alpha$ as defined above as an
effective running coupling constant is only justified at small $R$ where the
perturbative expression dominates.
  Also shown  are  the
two-loop perturbative results for $\alpha(R)$ for
different values of $\Lambda_R $.

\begin{figure}[htb]
\centering
\vspace{5cm}
\includegraphics{figsu3.ps}  % was -20
\caption{
The effective running coupling constant $\alpha(R)$ obtained from
the force between static quarks at separation $R$.  The scale is set
 by the string tension $K$. Data at $\beta=6.5$ are from
table~\protect\ref{alphat} (diamonds) and at $\beta=6.2$ (triangles).
 The dotted error bars represent an estimate
of the systematic error due to lattice artefact correction as described
in the text.  The curves are the two-loop perturbative expression with
$a(6.5)\Lambda_R=0.060$ (dotted) and 0.070 (continuous).
}
\end{figure}

The figure clearly shows a {\it running} coupling constant.  Moreover
the result is consistent with the expected perturbative dependence on
$R$ at small $R$.  There are systematic errors, however. At larger $R$,
the perturbative two-loop expression will not be an accurate estimate of
the measured potentials, while at smaller $R$, the lattice artefact
corrections are relatively big.  Setting the scale using $\sqrt K=0.44$
GeV implies $1/a(\beta=6.5)=4.13 $ GeV, so $R < 4a(6.5)$ corresponds to values
of $1/R > 1$ GeV.  This $R$-region is expected to be adequately
described by perturbation theory.  Another indication that perturbation
theory is accurate at such $R$-values is that $\Delta V_c / \Delta R$ at
small $R$ is found to be very much greater than the non-perturbative
value $K$ at large $R$.

Even though the lattice artefact correction of all 6 off-axis points by
one parameter is very encouraging, the only way to be certain that
lattice artefacts are eliminated is by the comparison of different
$\beta$ values (with thus different $R/a$ values at the same physical
$R$ value). Now this test was satisfied in an $SU(2)$
study~\cite{cmlms,ukqcd}.  Even so we can check independently in $SU(3)$
and we use UKQCD data \cite{uksu3} from $24^3 \times 48$ lattices
at $\beta =6.2$. We follow the
same procedure as above (for more details see~\cite{uksu3}) and obtain
for the string tension  $Ka^2= 0.0251(8)$. This
corresponds to a ratio of lattice spacings $a(6.2)/a(6.5)=1.484(35)$ to
be compared with the two-loop perturbative ratio of 1.404.
{}From the small $R$ results we then obtain a corrected force and
hence a running coupling. Setting the
scale from the measured string tensions, we also show the $\beta =6.2$
results for this effective running coupling in the figure. There is
excellent agreement with the results from $\beta=6.5$.

The easiest way to describe the value of the running coupling constant
$\alpha$
is in terms of a scale or $\Lambda$ value with the understanding that
we are only determining $\alpha$ for a range of energy scales
 $1/R$ - namely 1 to 3 GeV.
The final estimate of $\Lambda$ is made from the figure, weighting
smaller $R$ more heavily since the perturbative expression is
more accurate as $\alpha(R)$ becomes smaller. We exclude the lowest
$R$ point since the lattice artefact correction is for $R>a$.
  Remembering that the systematic errors due
to lattice artefact correction are estimates only and since these systematic
errors are dominant, we do not attempt a fit but we can  conclude
that our results are consistent with values of
 $\Lambda$ lying in the range shown by the two curves plotted.
{}From the data at $\beta $ = 6.5, these curves have
$a(6.5)\Lambda_R$=0.070 and 0.060.  Using the value of the string
tension from the fit, we get $\sqrt K/\Lambda_L$= 49.6(3.8).  Moreover, this
value is consistent with the evaluation at both $\beta=$ 6.5 and 6.2.
A rather similar analysis by Bali and Schilling~\cite{bali} also agrees
with this result.

\section{Conclusions}

Using  the bare
coupling $g$ derived from $\beta =6/g^{2}$ and the  two-loop  perturbative
 relationship  $a(g)$ in terms of the scale $\Lambda_L$
gave \cite{su3,uksu3,bali} the following
slowly decreasing values of $\sqrt K/\Lambda_L =$ 93.0(7) and
96.7(1.6)(2.6) at $\beta=6.0$; 85.9(1.5) and 86.4(1.0)(1.9) at $\beta=
6.2$ and
82.3(8)(1.7)
 at $\beta = 6.4$.  Our present analysis
at $\beta=6.5$ yields $\sqrt K/\Lambda_L =80.0(1.4)$. Clearly, the
$\beta \rightarrow \infty $ limit lies below these values. Moreover
the statistically significant decrease of these values is evidence that
two-loop perturbative scaling is not obtained in terms of the bare coupling.
Our present method which does not rely on the bare coupling
gives  the scaling result which should be independent of $\beta$.
Our estimate is  $ \sqrt K /\Lambda_L = 49.6(3.8)$.    This
is sufficiently far below the values extracted from the bare coupling
to imply  that asymptotic scaling to two-loop perturbation theory is
 not ``just around the corner'' but will only  be
satisfied accurately at larger $\beta$-values than those currently
accessible to lattice simulation.

Using an empirical definition of an effective coupling constant in
terms of the measured plaquette action, it is possible to extrapolate
$\sqrt K /\Lambda_L$ to $a=0$ and the resulting estimates~\cite{fhk,bali}
are in agreement with our value. This provides additional support
for the viewpoint that the bare lattice coupling has bad perturbative
behaviour but that perturbative descriptions of small distance
phenomena at existing $\beta$ values are reliable if a physical
coupling definition is used.

Our result for the running coupling $\alpha_R(R)$given in the figure and
table~\ref{alphat} can be read directly as $\alpha_{\msbar}(q)$ with
$q=1/R$ since these schemes are so close to each other. Since we obtain
results consistent with the perturbative evolution, we can estimate the
continuum ratio $\sqrt K / \Lambda_{\msbar}$ = 1.72(13) for pure SU(3)
theory. Setting the scale using $\sqrt K$ = 0.44 GeV, then gives
$\Lambda_{\msbar}$=256(20) MeV.  These results are obtained for rather
modest energies ( $1/R \approx 1$--$3$ GeV ) but there is evidence from
studies in $SU(2)$ where higher energies have been reached~\cite{ukqcd}
that the method is stable as the energy scale is increased somewhat.
{}From lattice results for ratios of other non-perturbative quantities
(glueball masses, critical temperature, etc) to the string tension, one
can then determine their value in terms of $\Lambda_{\msbar}$ as well.

 Even though the scales probed in this work are relatively small
(i.e. only 3 GeV), the agreement with the perturbative evolution
of the coupling constant implies that this is a reasonable way to
determine the coupling constant in terms of non-perturbative
physical quantities. We are able to determine $\Lambda_{\msbar}$
relatively accurately compared to experiment. Of course experiment
has full QCD with dynamical light quarks included while precise
lattice  simulation of full QCD is still a considerable challenge.


\begin{thebibliography}{99}

\bibitem{bill}
A. Billoire, Phys.\ Lett.\ 104B (1981) 472.

\bibitem{cmlms}
C.~Michael, Phys.\ Lett.\ 283B (1992) 103.

\bibitem{ukqcd}
S.P.~Booth et al.\ (the UKQCD Collaboration), Phys.\ Lett.\ B 275
(1992) 424; Liverpool preprint LTH284.

\bibitem{lm}
P. Lepage and P. Mackenzie, Nucl.\ Phys.\ B (Proc.\ Suppl.) 20 (1991) 173.

\bibitem{lu}
M. L\"uscher, P. Weisz and U. Wolff, Nucl.\ Phys.\ B359 (1991) 221;\\
M. L\"uscher, R. Sommer, U. Wolff and P. Weisz , CERN preprint
TH 6566/92, DESY preprint 92-096.

\bibitem{lth285}
S. P. Booth et al.\ (the UKQCD Collaboration), Liverpool preprint
LTH285; Phys. Lett. B (in press).

\bibitem{uksu3}
C. R. Allton et al.\ (the UKQCD Collaboration), Phys.\ Lett.\ B 284 (1992)
377 and in preparation.

\bibitem{mt}
C. Michael and M. Teper, Nucl.\ Phys.\ B 314 (1989) 347.

\bibitem{bali}
G. Bali and K. Schilling, Wuppertal preprints WUB 92-02; WUB 92-29 (1992).

\bibitem{su3}
S. Perantonis and C. Michael, Nucl.\ Phys.\ B 347 (1990) 854.

\bibitem{fhk}
J. Fingberg, U. Heller and F. Kasrch, Bielefeld preprint BI-TP 92-26.

\end{thebibliography}
\end{document}